# Greenberger-Horne-Zeilinger-type violation of local realism by mixed states

X.-Q. Zhou,[1] C.-Y. Lu,[1] W.-B. Gao,[1] J. Zhang,[1] Z.-B. Chen,[1] T. Yang,[1] and J.-W. Pan[1,2]

[1]*Hefei National Laboratory for Physical Sciences at Microscale and Department of Modern Physics, University of Science and Technology of China, Hefei, 230026, People's Republic of China*
[2]*Physikalisches Institut, Universität Heidelberg, Philosophenweg 12, D-69120 Heidelberg, Germany*
(Dated: May 23, 2008)

Cluster states are multi-particle entangled states with special entanglement properties particularly suitable for quantum computation. It has been shown that cluster states can exhibit Greenberger-Horne-Zeilinger (GHZ)-type non-locality even when some of their qubits have been lost. In the present work, we generated a four-photon mixed state, which is equivalent to the partial, qubit-loss state of an N-qubit cluster state up to some local transformations. By using this mixed state, we then realize a GHZ-type violation of local realism. Our results not only demonstrate a mixed state's GHZ-type non-locality but also exhibit the robustness of cluster states under qubit-loss conditions.

Multi-particle entangled states play a fundamental role in the field of quantum information and its applications. In recent years, a class of multi-particle entangled states, cluster states, have attracted considerable interest, mainly due to their applications in the "one-way" model for universal quantum computation [1]. In that model, one can perform quantum computing simply by using single-qubit measurements and feed-forward instead of unitary evolution.

Besides their fascinating use in quantum computation, cluster states are a novel kind of multi-particle entangled states with fundamentally new and different properties. When qubit losses occur, unlike Greenberger-Horne-Zeilinger (GHZ) states which would be totally disentangled, cluster states can still maintain some entanglement [1]. A recent theoretical paper [2] shows that cluster states can be used to construct a GHZ-type argument to refute local realism (LR), as experimentally tested by Walther *et al.* [3] and Kiesel *et al.* [4] using four-photon cluster states. Moreover, such a refutation is applicable even under qubit-loss conditions, i.e., for any partial, hence mixed, state of a small number of connected qubits [a five-qubit partial state in the case of an N-qubit ($N > 5$) linear cluster]. This is very different from other GHZ-type non-locality arguments [5–10] which are all based on pure states. In the present paper, we generated a four-photon mixed state, which in an ideal case is equivalent, up to some local transformations, to the qubit-loss state of an $N$-qubit cluster state. By using this mixed state, we then realize a GHZ-type violation of LR and thus exhibit the robustness of cluster states under qubit-loss condition.

First, let us see how to use an $N$-qubit cluster state to construct the GHZ-type argument. Instead of a linear cluster as considered in Ref. [2], we consider a T-shaped $N$-qubit cluster state $|\phi_N\rangle$ shown in Fig. 1. Based on the definition of cluster states, it can be characterized by a set of eigen-equations:

$$\sigma_X^{(1)}\sigma_Z^{(2)}\sigma_0^{(3)}\sigma_0^{(4)}\sigma_0^{(5)}...\sigma_0^{(N-1)}\sigma_0^{(N)}|\phi_N\rangle = |\phi_N\rangle, (E_1)$$
$$\sigma_Z^{(1)}\sigma_X^{(2)}\sigma_Z^{(3)}\sigma_Z^{(4)}\sigma_0^{(5)}...\sigma_0^{(N-1)}\sigma_0^{(N)}|\phi_N\rangle = |\phi_N\rangle, (E_2)$$
$$\sigma_0^{(1)}\sigma_Z^{(2)}\sigma_X^{(3)}\sigma_0^{(4)}\sigma_0^{(5)}...\sigma_0^{(N-1)}\sigma_0^{(N)}|\phi_N\rangle = |\phi_N\rangle, (E_3)$$
$$...\qquad(1)$$
$$\sigma_0^{(1)}\sigma_0^{(2)}\sigma_0^{(3)}\sigma_0^{(4)}\sigma_0^{(5)}...\sigma_Z^{(N-1)}\sigma_X^{(N)}|\phi_N\rangle = |\phi_N\rangle, (E_N)$$

where $\sigma_i^{(j)}$ denote Pauli matrix $\sigma_i$ on qubit $j$. By multiplication using the algebra of Pauli matrices, we can derive the following three equations:

$$C_1 = E_1E_2 : \sigma_Y^{(1)}\sigma_Y^{(2)}\sigma_Z^{(3)}\sigma_Z^{(4)}\sigma_0^{(5)}...\sigma_0^{(N-1)}\sigma_0^{(N)}|\phi_N\rangle = +|\phi_N\rangle,$$
$$C_2 = E_2E_3 : \sigma_Z^{(1)}\sigma_Y^{(2)}\sigma_Y^{(3)}\sigma_Z^{(4)}\sigma_0^{(5)}...\sigma_0^{(N-1)}\sigma_0^{(N)}|\phi_N\rangle = +|\phi_N\rangle,$$
$$C_3 = E_1E_2E_3 : \sigma_Y^{(1)}\sigma_X^{(2)}\sigma_Y^{(3)}\sigma_Z^{(4)}\sigma_0^{(5)}...\sigma_0^{(N-1)}\sigma_0^{(N)}|\phi_N\rangle = -|\phi_N\rangle.$$
$$(2)$$

Now let us focus on equations $E_2$, $C_1$, $C_2$, and $C_3$. From the point of view of LR, if the measurements corresponding to the four operators are mutually spacelike performed on the cluster state ($\sigma_0^{(i)}$ means do nothing on qubit $i$), four equations $Z_1X_2Z_3Z_4 = +1$, $Y_1Y_2Z_3Z_4 = +1$, $Z_1Y_2Y_3Z_4 = +1$, $Y_1X_2Y_3Z_4 = -1$, can be deduced from the fully-correlated results, where $X_j$ ($Y_j$, $Z_j$) represents element of reality with value $+1/-1$ for D/A (R/L, H/V) polarizations [i.e., corresponding to $\sigma_X^{(j)}$ ($\sigma_Y^{(j)}$, $\sigma_Z^{(j)}$)]. By multiplying the above four equations, one can then easily obtain the paradox $(Y_1Z_1X_2Y_2Y_3Z_3Z_4Z_4)^2 = -1$, which demonstrates that the theory of LR is self-contradictory.

From the above GHZ-type argument, one can find that all the measurements only involve qubits 1-4, which means that no matter what happens to the other N-4 qubits in the cluster, even all of them have been lost, it will make no difference to the GHZ-type argument. In other word, the partial state containing only qubits 1-4 of the cluster, which is a mixed state, is entangled enough to exhibit the GHZ-type non-locality. Note that we need a five-qubit partial state in the case of a linear cluster [2].

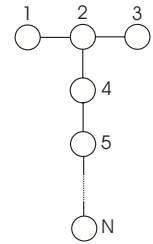

FIG. 1: A T-shaped N-qubit cluster state

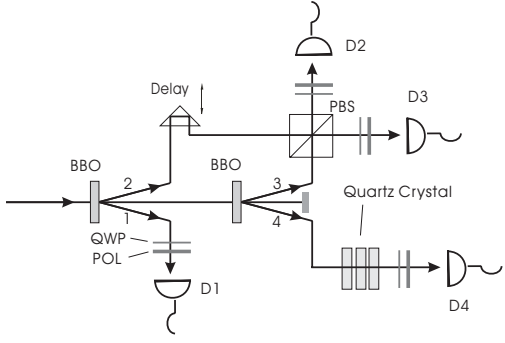

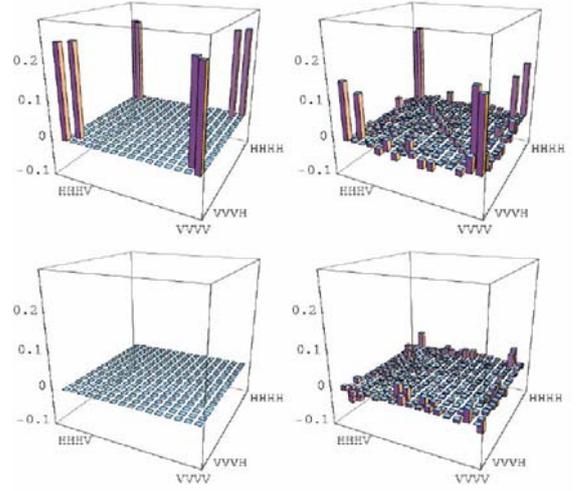

FIG. 2: Experimental setup for generating $\rho_\psi$. Two pairs of entangled photons are produced after an ultraviolet (UV) laser passing through two BBO crystals. The UV laser has a central wavelength of 394 nm, a pulse duration of 120 fs, a repetition rate of 76 MHz, and an average pump power of ~900 mW. Without quartz crystals, coincidences between detectors D1, D2, D3, and D4 would exhibit a four-photon GHZ entanglement. Three pieces of quartz crystals (1cm width), whose fast axis are all oriented at 45°, convert the GHZ state into our desired mixed state $\rho_\psi$. Polarizers (POL) and quarter waveplates (QWP) before the four detectors are used to perform the measurements of $H/V$, $D/A$, or $R/L$ polarization.

FIG. 3: Density matrix of the four-photon mixed state in the H/V basis. Shown are the real (top) and imginary (bottom) parts of the density matrix for the ideal case (left) and the reconstruction from the experimental four-photon tomography data (right). The real density matrix was reconstructed by way of a maximum likelihood method using four-photon coincidence rates obtained in 256 polarization projections.

By easy calculation, one can deduce that this partial state $\rho_\phi$ containing qubits 1-4 can be written in the form $\rho_\phi = \frac{1}{2}(|\phi_1\rangle\langle\phi_1| + |\phi_2\rangle\langle\phi_2|)$, where $|\phi_1\rangle = \frac{1}{\sqrt{2}}(|DHD\rangle_{123} + |AVA\rangle_{123}) \otimes |H\rangle_4$ and $|\phi_2\rangle = \frac{1}{\sqrt{2}}(|DHD\rangle_{123} - |AVA\rangle_{123}) \otimes |V\rangle_4$. For experimental convenience, we will use a slightly different four-qubit mixed state $\rho_\psi$ to do the non-locality test. It is of the form $\rho_\psi = \frac{1}{2}(|\psi_1\rangle\langle\psi_1| + |\psi_2\rangle\langle\psi_2|)$, where $|\psi_1\rangle = \frac{1}{\sqrt{2}}(|HHH\rangle_{123} + |VVV\rangle_{123}) \otimes |D\rangle_4$ and $|\psi_2\rangle = \frac{1}{\sqrt{2}}(|HHH\rangle_{123} - |VVV\rangle_{123}) \otimes |A\rangle_4$. Since $\rho_\phi$ and $\rho_\psi$ are equivalent only up to some local transformations, following the same reasoning, one can easily obtain four equations, $X_1X_2X_3X_4 = +1$, $X_1Y_2Y_3X_4 = -1$, $Y_1X_2Y_3X_4 = -1$ and $Y_1Y_2X_3X_4 = -1$, to deduce GHZ-type violation of LR.

The experimental setup to generate the mixed state $\rho_\psi$ is shown in Fig. 2. First, we create a four-photon GHZ state using the standard method as in Refs. [11, 12]. UV laser pulses from frequency-doubled Ti:Sapphire laser pass through two $\beta$-barium borate (BBO) crystals [13], with walk-off compensation to produce two photon pairs, each in the Bell state $|\Phi^+\rangle = \frac{1}{\sqrt{2}}(|HH\rangle + |VV\rangle)$. Then photons 2 and 3 are steered to a polarizing beam splitter (PBS) where the path length of each photon has been adjusted such that they arrive simultaneously. Since the PBS transmits $H$ and reflects $V$ polarization, coincidence detection between the two outputs of PBS implies that both photons 2 and 3 are either $H$-polarized or $V$-polarized, and thus projects the four-photon state onto a two-dimensional subspace spanned by $HHHH$ and $VVVV$.

After the PBS, the renormalized state corresponding to a fourfold coincidence is thus

$$|\Psi_{GHZ}\rangle = \frac{1}{\sqrt{2}}(|HHHH\rangle_{1234} + |VVVV\rangle_{1234}), \quad (3)$$

which exhibits four-photon GHZ entanglement. The GHZ state can be written in the form

$$|\Psi_{GHZ}\rangle = \frac{1}{\sqrt{2}}\left(\left|GHZ^+\right\rangle_{123} \otimes |D\rangle_4 + \left|GHZ^-\right\rangle_{123} \otimes |A\rangle_4\right), \quad (4)$$

where $|GHZ^\pm\rangle_{123} = \frac{1}{\sqrt{2}}(|HHH\rangle_{123} \pm |VVV\rangle_{123})$.

Then, three pieces of quartz crystals (1 cm width) with their fast axes at 45° are placed into the path of photon 4 to decohere the state. Due to the birefringence of the quartz crystals, the $|D\rangle$ component of photon 4 moves faster than the $|A\rangle$ component. The optical path difference of these two components is larger than the coherence length of photon 4, and as such the two components $|D\rangle$ and $|A\rangle$ are no longer coherent. Thus, the four-photon GHZ state has been converted into the desired mixed state $\rho_\psi$.

Then, we use the method of quantum state tomography to fully characterize our state from a discrete set of measurements. For a four-photon polarization state, like our mixed state $\rho_\psi$, the whole density matrix $\rho$ is a $16 \times 16$-dimensional object that can be reconstructed by linear combinations of 256 linearly independent four-photon polarization projections. We perform each of these 256 correlation measurements for 60 s using all combinations of {$H, V, D, R$}. A maximum of 182 fourfold coincidence counts in 60 s are measured in the case of the setting $VVVD$. Instead of a direct linear combination of measurement results, which can lead to unphysical density matrices owing to experimental noise, we use a maximum-likelihood reconstruction technique [14–16]. The measured density matrix $\rho_{meas}$ is shown in Fig. 2. Our reconstructed state is in good agreement with the target mixed state. This can be quantified by the state fidelity $F = |Tr\left(\sqrt{\sqrt{\rho_\psi}\rho_{meas}\sqrt{\rho_\psi}}\right)|^2 = 0.68 \pm 0.02$, which is the overlap between $\rho_{meas}$ and the ideal state $\rho_\psi$. Note that any pure state cannot have a fidelity greater than 0.5 with our target mixed state.

To futher ensure that the generated state is a mixture of two three-party GHZ-entangled states, we apply entanglement witnesses for the two components of $\rho_\psi$. We observe that, when photon 4 is on the state $|D\rangle$ ($|A\rangle$), the fidelity of the other three-photon state with respect to $|GHZ^+\rangle_{123}$ ($|GHZ^-\rangle_{123}$) is $0.74 \pm 0.01$ ($0.72 \pm 0.01$), yielding expectation values of the entanglement witness of $\langle W_{GHZ^+_{123}}\rangle \equiv \langle \frac{I}{2} - |GHZ^+_{123}\rangle\langle GHZ^+_{123}|\rangle = -0.24 \pm 0.01$ ($\langle W_{GHZ^-_{123}}\rangle \equiv \langle \frac{I}{2} - |GHZ^-_{123}\rangle\langle GHZ^-_{123}|\rangle = -0.22 \pm 0.01$), clearly proving the genuine three-photon entanglement [17] for both of the two components in $\rho_\psi$.

There are three main reasons for the imperfection of our state. First, high-order emissions of entangled photons give rise to the undesired components [18]. Second, the partial distinguishability of independent photons results in the quality reduction of the initial GHZ state, which eventually reduce the fidelity of our state. Third, the imperfection of the decoherence process further reduce the quality of the state. Before decoherence was applied, the fidelity of the initial state with four-qubit GHZ state is $0.78 \pm 0.02$.

Demonstration of the conflict between LR and QM in our case consists of four measurements. First, we perform $XYYX$, $YXYX$ and $YYXX$ measurements. If the results obtained are in agreement with the predictions for the mixed state $\rho_\psi$, then for an $XXXX$ measurement, our consequent expectations using LR are exactly the opposite of our expectations using QM.

For each measurement, we have 16 possible outcomes, 8 of which should never occur in an ideal case. We will follow two independent possible strategies to explain our experimental results. In the first strategy we simply assume that the spurious events are attributable to experimental imperfection that is not correlated to the elements of reality a photon carries. A local realist might argue against that approach and suggest that the non-perfect detection events indicate that the GHZ argument is inapplicable. In our second strategy we therefore accommodate local-realist theories, by assuming that the non-perfect events in the first three measurements indicate a set of elements of reality which are in conflict with quantum mechanics. We then compare the local realisic prediction for the *XXXX* experiment obtained under that assumption with the experimental results.

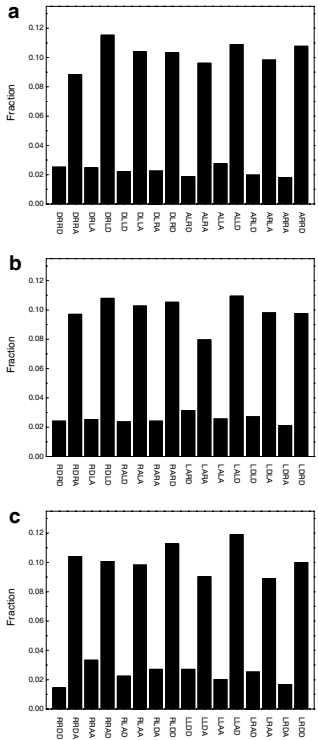

FIG. 4: All outcomes observed in the $XYYX$, $YXYX$, and $YYXX$ measurements. The experimental data show that we observe the terms predicted by QM (tall bar) in a fraction of $0.822 \pm 0.009$, $0.798 \pm 0.009$, $0.812 \pm 0.009$ of each cases.

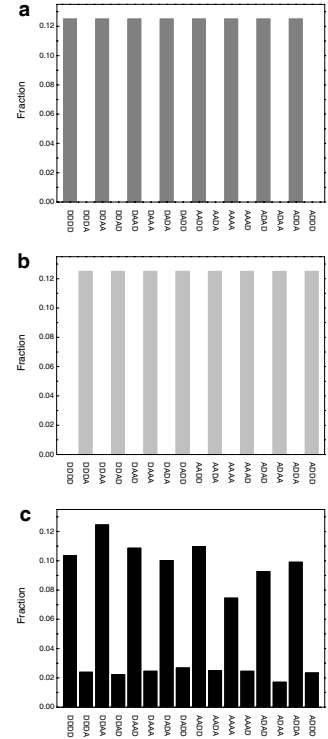

FIG. 5: Predictions of QM (a) and of LR (b), and observed results (c) for the $XXXX$ experiment. The experimental results clearly confirm the quantum predictions within experimental error and are in conflict with LR.

For *XYYX*, *YXYX* and *YYXX* measurements, the actually measured results have 48 possible outcomes whose individual fractions are shown in Fig. 4. Adopting our first strategy, we assume that the spurious events are attributed to unavoidable experimental errors. Within the experimental accuracy, we conclude that the desired correlations in these experiments confirm the quantum predictions for $\rho_\psi$. Then, we compare the predictions of QM and LR with the results of an *XXXX* measurement (Fig. 5). Again within experimental error, the fourfold coincidences predicted by QM, but not by LR, occur. In this sense, we believe that we have experimentally realized a GHZ-type violation of LR using a mixed state.

We then investigate whether LR could reproduce the experimental results shown in Fig. 5, if we assume that the spurious non-GHZ events in the other three measurements (Fig. 4) actually indicate a deviation from quantum physics. To answer this we adopt our second strategy and consider the best prediction a local realistic theory could obtain using these spurious terms. How, for example, could a local realist obtain the quantum prediction $|DDDD\rangle_{1234}$. One possibility is to assume that triple events producing $|DDDD\rangle_{1234}$ would be described by a specific set of local hidden variables such that they would give events that are in agreement with quantum theory in both an *XYYX* and a *YXYX* experiment (for example, the results $|DLRD\rangle_{1234}$ and $|LDRD\rangle_{1234}$), but give a spurious event for a *YYXX* experiment (in this case, $|LLDD\rangle_{1234}$). In this way any local realistic prediction for an event predicted by quantum theory in our *XXXX* measurement will use at least one spurious event in the earlier measurements together with two correct ones. Therefore, the fraction of correct events in the *XXXX* measurement can at most be equal to the sum of the fractions of all spurious events in the *XYYX*, *YXYX*, and *YYXX* measurements, that is, $0.57 \pm 0.016$. However, we experimentally observed such terms with a fraction of $0.81 \pm 0.009$ (Fig. 5), which violates the local realistic expectation by more than 12 standard deviations.

Our latter argument is equivalent to adopting a Mermin-type inequality [19]. For any LR model one has $\langle S \rangle_{LRT} \leq 2$, where $S = XXXX - XYYX - YXYX - YYXX$. We obtained positive expectation value of $0.626 \pm 0.019$ for *XXXX* measurement and negative expectation values of $-0.646 \pm 0.018$, $-0.595 \pm 0.018$, and $-0.628 \pm 0.019$ for *XYYX*, *YXYX*, and *YYXX* measurements, respectively. The observed value for $S$ is $2.50 \pm 0.04$, which is a violation by about $12\sigma$.

Cluster states have special entanglement properties such as strong entanglement persistency which GHZ-type entangled states do not have. Such features lead to the surprising non-locality properties shown here and may relate to some interesting implications in loss-tolerant one-way quantum computing etc. [20]. Former experiments[3, 4, 21] have shown that cluster states can still exhibit some entanglement after qubit loss occurs. In this work, we further demonstrated the strong entanglement persistency of cluster states by GHZ-type violation of LR with qubit-loss cluster states. It is also, to our knowledge, the first realization of GHZ-type violation of LR by a mixed state. Of course, as in almost all of the existing experiments testing LR, our experiment also has certain well-known loopholes, such as the locality and efficiency loopholes. We hope these loopholes can be closed in future experiments.

We thank V. Scarani, M. Zukowski and X.-H. Bao for helpful discussions. This work was supported by the NNSF of China, the CAS, the National Fundamental Research Program (under Grant No. 2006CB921900) and the Fok Ying Tung Education Foundation.

[1] H.J. Briegel and R. Raussendorf, Phys. Rev. Lett. **86**, 910 (2001); R. Raussendorf and H.J. Briegel, Phys. Rev. Lett. **86**, 5188 (2001); P. Walther *et al.*, Nature (London) **434**, 169 (2005); R. Prevedel *et al.*, Nature **445**, 65 (2007).
[2] V. Scarani, A. Acín, E. Schenck, and M. Aspelmeyer, Phys. Rev. A **71** 042325 (2005).
[3] P. Walther, M. Aspelmeyer, K.J. Resch, and A. Zeilinger, Phys. Rev. Lett. **95**, 020403 (2005).
[4] N. Kiesel, C. Schmid, U. Weber, G. Tóth, O. Gühne, and H. Weinfurter, Phys. Rev. Lett. **95**, 210502 (2005)
[5] D.M. Greenberger, M.A. Horne, A. Shimony, and A. Zeilinger, Am. J. Phys. **58**, 1131 (1990).
[6] W. Tittel, J. Brendel, H. Zbinden, and N. Gisin, Phys. Rev. Lett. **81**, 3563 (1998); A. Stefanov, H. Zbinden, N. Gisin, and A. Suarez, Phys. Rev. Lett. **88**, 120404 (2002).
[7] J.-W. Pan, D. Bouwmeester, M. Daniell, H. Weinfurter, and A. Zeilinger, Nature (London) **403**, 515 (2000).
[8] A. Cabello, Phys. Rev. Lett. **86**, 1911 (2001); *ibid.* **87**, 010403 (2001); *ibid.* **95**, 210401 (2005).
[9] Z.-B. Chen *et al.*, Phys. Rev. Lett. **90**, 160408 (2003).
[10] C. Cinelli *et al.*, Phys. Rev. Lett. **95**, 240405 (2005); T. Yang *et al.*, *ibid.* **95**, 240406 (2005).
[11] A. Zeilinger, M. Horne, H. Weinfurter, and M. Zukowski, Phys. Rev. Lett. **78**, 3031 (1997).
[12] J.-W. Pan, M. Daniell, S. Gasparoni, G. Weihs, and A. Zeilinger, Phys. Rev. Lett. **86**, 4435 (2001).
[13] P. Kwiat *et al.*, Phys. Rev. Lett. **75**, 4337 (1995).
[14] A. White, D. James, P. Eberhard, and P. Kwiat, Phys. Rev. Lett. **83**, 3103 (1999).
[15] D. James, P. Kwiat, W. Munro, and A. White, Phys. Rev. A **64**, 052312 (2001).
[16] Matlab codes for this method are from P.G. Kwiat's group: *http://research.physics.uiuc.edu/QI/Photonics/Tomography/*
[17] M. Bourennane *et al.*, Phys. Rev. Lett. **92**, 087902 (2004).
[18] V. Scarani, *et al.*, Eur. Phys. J. D **32**, 129 (2005); M. Barbieri, Phys. Rev. A **76**, 043825 (2007).
[19] N.D. Mermin, Phys. Today **43**, 9 (1990).
[20] M. Varnava, D. Browne, T. Rudolph, Phys. Rev. Lett. **97**, 120501 (2006); M. Varnava, D. Browne, T. Rudolph, Phys. Rev. Lett. **100**, 060502 (2008).
[21] C. Schmid, N. Kiesel, W. Wieczorek, and H. Weinfurter, New J. Phys. **9**, 236 (2007).